\documentclass[aps,reprint]{revtex4-1}
\usepackage{amsmath,amssymb}
\usepackage{bm}
\usepackage{graphicx}

\begin{document}
\title{Mechanics of cell crawling by means of force-free cyclic motion}%
\author{Mitsusuke Tarama}%
\email{tarama@fukui.kyoto-u.ac.jp}%
\affiliation{
Fukui Institute for Fundamental Chemistry, Kyoto University, Kyoto 606-8103, Japan
}%
\author{Ryoichi Yamamoto}%
\email{ryoichi@cheme.kyoto-u.ac.jp}%
\affiliation{
Department of Chemical Engineering, Kyoto University, Kyoto 615-8510, Japan
}%
\affiliation{
Institute of Industrial Science, The University of Tokyo, Tokyo 153-8505, Japan
}%

\date{\today}%

\begin{abstract}
The mechanics of crawling cells on a substrate is investigated by using a minimal model that satisfies the force-free condition. 
A cell is described by two subcellular elements connected by a linear actuator that changes the length of the cell cyclically in time, together with periodic alternation of adhesive characters at the interface between the cell and the substrate. 
Here the key model parameters are the phase shifts between the elongation of the actuator and the alternation of the adhesion of the two elements. 
We emphasize that the phase shifts determine not only the efficiency of the crawling motion but also its direction. 
\end{abstract}
\maketitle

Dynamics of active particles have attracted much attention from physicists over the past decade. 
In contrast to passive particles, active particles exhibit spontaneous motion such as directional translation without external forcing. 
This property of a vanishing force monopole is known as the force-free condition. 
Therefore, it is not trivial how active particles such as microorganisms can achieve a net translational motion from an internal cyclic motion. 

In his pioneering lecture, Purcell shed light on the importance of breaking the time-reversal symmetry~\cite{Purcell1977Life}. 
His idea was theoretically investigated later by using a simple model swimmer composed of three linked spheres at low Reynolds number~\cite{Najafi2004Simple}. 
The three-bead swimmer with autonomous oscillation of the bond lengths was also studied by considering viscoelastic bonds connecting the beads~\cite{Gunther2008A}. 
Recent studies extend the idea to three-bead swimmers in a viscoelastic solvent~\cite{Yasuda2017Swimmer-Microrheology,Yasuda2017Elastic}. 
The time-reversal symmetry can be broken by a phase shift between the periodic oscillations of at least two active linkers~\cite{Golestanian2008Analytic}. 
All these studies are concerning microswimmers, i.e., microscopic objects that are moving in a fluid environment.

In contrast, there also exist microorganisms that migrate on substrates~\cite{Fletcher2004An}. 
Such crawling motion is observed in many Eukaryotic cells, including \textit{keratocyte} and \textit{Dictyostelium} cells. 
Typically, the mechanism of the crawling motion of biological cells is widely believed to consist of the following four steps~\cite{Ananthakrishnan2007The}: 
1) Protrusion of the leading edge, 2) adhesion of the leading edge, 3) deadhesion at the trailing edge, and 4) contraction of the trailing edge. 
A number of studies have been conducted to understand the underlying physics of each step, especially the protrusion due to the actin polymerization~\cite{Mogilner2009The,Doubrovinski2011Cell,Blanchoin2014Actin} and the contraction of actomyosin~\cite{Julicher2007Active,Prost2015Active,Hawkins2011Spontaneous}. 
Astonishingly, however, to our knowledge, this cycle of the crawling mechanism itself has never been verified systematically.

The purpose of this letter is to investigate the basic mechanics of a crawling cell with a focus on the cycle of the protrusion, the contraction, and the adhesion to the substrate. 
\begin{figure}[b]
\begin{center}
 \includegraphics[width=\columnwidth]{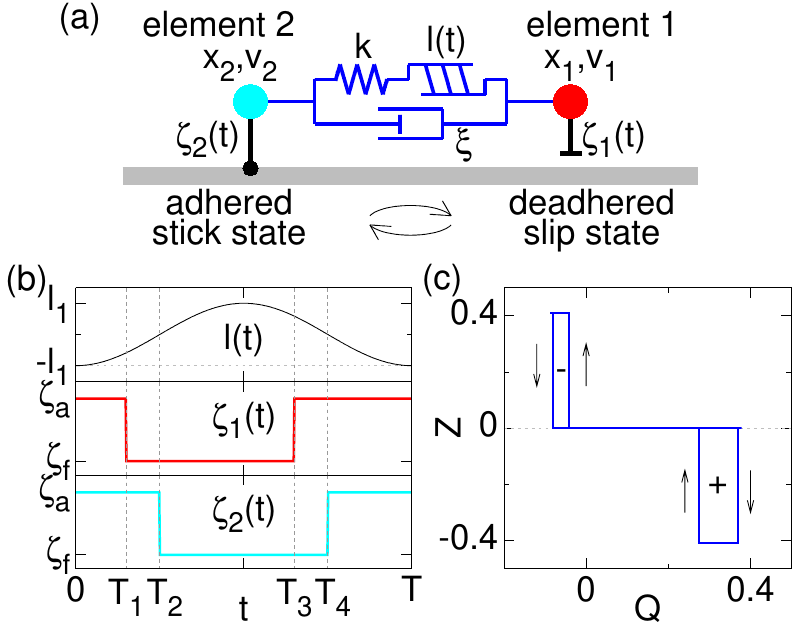}
\caption{(Colour online)
(a) Schematics of the two-element model of a cell crawling on a substrate (grey). 
The red and cyan elements are connected by a viscoelastic bond (blue), which consists of a spring of elasticity $k$, a dashpot with dissipation rate $\xi$, and a linear actuator with time-dependent length $\ell(t)$. 
The substrate friction $\zeta_i(t)$ of each element switches between the stick and slip states. 
(b) An example of the time series for $\ell(t)$, $\zeta_1(t)$, and $\zeta_2(t)$, and (c) the corresponding trajectory in the $Q$--$Z$ space, where the signed area enclosed by the curve denotes the distance over which the cell travels during the corresponding time interval. 
In panel (c), the black arrows show the time evolution direction and the plus and minus signs indicate the temporal forward and backward migration, respectively. 
}%
\label{fig:two-element_schematics}
\end{center}
\end{figure}
In order to make the system as simple as possible, we employ a minimum model that satisfies the force-free condition where a cell is described by two subcellular elements connected by a viscoelastic bond of Kelvin-Voigt type. 
We assume that the intracellular activities change the friction of the elements and the bond length, as sketched in fig.~\ref{fig:two-element_schematics}(a).

Since the typical size of a cell is of the order of ten micrometers, the inertia is negligible. 
Then, the force balance equation of each element is given by
\begin{align}
\zeta_1(t) v_1 +\xi ( v_1 -v_2) &= -k \{ x_{12} -(\ell_0 +\ell(t)) \},
 \label{eq:force_balance_1}\\
\zeta_2(t) v_2 +\xi ( v_2 -v_1 ) &= k \{ x_{12} -(\ell_0 +\ell(t)) \},
 \label{eq:force_balance_2}
\end{align}
where $x_i$ and $v_i$ are the position and the velocity of the element $i$, and $x_{12} = | x_2 -x_1 |$ is their distance. 
The first terms on the left-hand side of eqs.~\eqref{eq:force_balance_1} and \eqref{eq:force_balance_2} represent the substrate friction, which we assume is a simple linear function of the velocity with the coefficient $\zeta_i(t)$. 
The second terms stand for the dashpot describing the intracellular dissipation with rate $\xi$. 
The intracellular elasticity of the cell is taken into account by the harmonic spring with strength $k$ and free length $\ell_0$, which is connected in series by a linear actuator of length $\ell(t)$. 
We emphasise that eqs.~\eqref{eq:force_balance_1} and \eqref{eq:force_balance_2} satisfy the force-free condition. 

In actual cells, protrusion and contraction, as well as adhesion to and deadhesion from the substrate, occur as a result of complicated intracellular chemical reactions. 
Instead of introducing explicitly such intracellular activities, here we simply regard the protrusion due to the actin polymerisation and the actomyosin contraction as a cyclic elongation of the cell body. 
We naively include such cyclic elongation by a linear actuator that changes the length periodically: 
\begin{equation}
\ell(t) = -\ell_1 \cos \omega t,
 \label{eq:ell}
\end{equation}
where $\ell_1$ represents the magnitude of the actuator elongation and $\omega$ is its frequency. 
Here we assume a sinusoidal change with period $T = 2\pi /\omega$. 
The time $t$ is measured with respect to the time at which the length of the actuator takes its minimum value.

In addition, the adhesion-deadhesion transition between the cell and the substrate underneath is included by the change in the substrate friction coefficient $\zeta_i(t)$. 
Since it is often assumed to be a sharp transition~\cite{Barnhart2015Balance}, we consider that $\zeta_i(t)$ switches between the adhered stick state and the deadhered (free) slip state: 
\begin{equation}
\zeta_i(t) = \left\{
\begin{array}{ll}
\zeta_f & \textrm{if~} 2 m_i \pi < \omega t -\psi_i \le (2 m_i +1) \pi \\
\zeta_a & \textrm{if~} (2 m_i +1) \pi < \omega t -\psi_i \le 2 (m_i +1) \pi
\end{array}
 \right.
 \label{eq:zeta}
\end{equation}
where $m_i$ is an integer. 
$\zeta_a$ and $\zeta_f$ are the values of the friction coefficient during the stick and slip states, respectively. 
Note that the subscript f in $\zeta_f$ stands for ``free'', and represents the deadhered state. 
The frequency $\omega$ is chosen to be the same as that of the actuator elongation, for simplicity. 
The phase shift $\psi_i$ is measured with respect to the phase of the elongation $\ell(t)$. 
In general, $\psi_1$ and $\psi_2$ may differ, which are the two key parameters for the current model. 

In order to solve eqs.~\eqref{eq:force_balance_1} and \eqref{eq:force_balance_2} analytically, we introduce the centre-of-mass velocity $V(t) = ( v_1 + v_2 ) /2$ and the actual elongation of the bond $Q(t) = x_{12} -\ell_0$. 
Then, eqs.~\eqref{eq:force_balance_1} and \eqref{eq:force_balance_2} reduce to
\begin{gather}
\frac{d Q}{dt} = -\lambda ( Q -\ell(t) ),
 \label{eq:dQ/dt}\\
V = Z \frac{dQ}{dt},
 \label{eq:V_dQ/dt}
\end{gather}
where $Z =(\zeta_2-\zeta_1) / 2(\zeta_1+\zeta_2)$ and
\begin{equation}
\lambda = k \Big( \frac{\zeta_1 \zeta_2}{\zeta_1+\zeta_2} +\xi \Big)^{-1} 
 \label{eq:lambda}
\end{equation}
is the characteristic damping rate of the elongation $Q$. 
Note that both $\zeta_1$ and $\zeta_2$ are positive, and so is $\lambda$. 
Once eq.~\eqref{eq:dQ/dt} is solved for $Q(t)$, then, from eq.~\eqref{eq:V_dQ/dt}, the centre-of-mass velocity $V(t)$ is obtained. 
By integrating $V(t)$, we can calculate the displacement of the centre of mass of the cell. 

Now we consider the net displacement of the centre-of-mass position in one cycle:
$\Delta R = \int_0^T dt V(t)$.
To this end, we confine ourselves to steady-state solutions hereafter. 

Since $\zeta_i$ switches stepwise as defined in eq.~\eqref{eq:zeta}, we can find an interval $T_{n-1} < t < T_n$ where both $\zeta_1$ and $\zeta_2$ take constant values, which we denote $\zeta_1^{(n)}$ and $\zeta_2^{(n)}$, respectively. 
Within this interval, eq.~\eqref{eq:dQ/dt} is integrated straightforwardly as
\begin{equation}
Q(t) = Q(T_{n-1}) e^{-\lambda_n (t -T_{n-1})} +I_n(t) e^{-\lambda_n t},
 \label{eq:Q_solution}
\end{equation}
where $\lambda_n$ denotes the value of eq.~\eqref{eq:lambda} calculated with $\zeta_1^{(n)}$ and $\zeta_2^{(n)}$, and $I_n(t)$ is given by
\begin{align}
I_n(t) 
&= - \frac{\ell_1 \lambda_n}{ \omega^2 +\lambda_n^2 }
 \big[ 
 ( \lambda_n \cos[\omega t] +\omega \sin[\omega t] ) e^{\lambda_n t} \notag\\
 &- ( \lambda_n \cos[\omega T_{n-1}] +\omega \sin[\omega T_{n-1}] ) e^{\lambda_n T_{n-1}}
  \big]. 
 \label{eq:I_n_psi_solution}
\end{align}
Here, note that we can divide one period into four intervals by $T_1 = \psi_1 /\omega$, $T_2 = (\psi_2 +n' \pi) /\omega$, $T_3 = T_1 +T/2$, $T_4 = T_2 +T/2$, and $T_0 = T_4 -T$, where $n'$ is an integer that satisfies $T_1 \le T_2 < T_3$. 
See fig.~\ref{fig:two-element_schematics}(b). 

By using eqs.~\eqref{eq:V_dQ/dt} and \eqref{eq:Q_solution}, the displacement of the centre-of-mass position in one cycle is calculated as
\begin{equation}
\Delta R = \sum_{n=1}^4 Z_n ( Q_n -Q_{n-1} ),
 \label{eq:DeltaR_steady-state}
\end{equation}
where $Z_n$ is the value of $Z$ calculated for $\zeta_1^{(n)}$ and $\zeta_2^{(n)}$. 
We have written $Q_n = Q(T_n)$ for $n=1,2,3,4$ and $Q_0 = Q_4$. 
Note that, for a steady state, the elongation should be the same after one cycle. 
Equation~\eqref{eq:DeltaR_steady-state} denotes that the net displacement $\Delta R$ does not depend on $\ell_0$, while, from eqs.~\eqref{eq:Q_solution} and \eqref{eq:I_n_psi_solution}, it is proportional to $\ell_1$. 
It also indicates that $\Delta R$ is given by the signed area enclosed by the trajectory in the $Q$--$Z$ space as depicted in fig.~\ref{fig:two-element_schematics}(c). 

Then, the efficiency of the cell is calculated as follows. 
The input energy is evaluated by the work supplied by the actuator $W_{in}$. 
However, since the cell is force free and crawling horizontally on a substrate, it does not apply any actual work. 
Therefore we introduce an additional constant force on the cell $f_{\epsilon}$ which is sufficiently small so that it does not affect the dynamics. 
Then, the work done by the cell over one period is given by $f_{\epsilon} \Delta R$. 
Consequently, we define the efficiency as
\begin{equation}
\eta = \frac{\Delta R /W_{in}}{\max(\Delta R /W_{in})}.
 \label{eq:eta}
\end{equation}
Note that the efficiency is normalized by the maximum value, which eliminates the dependence on $f_{\epsilon}$. 
See the supplemental material~\cite{Supplement} for the explicit formula. 

\begin{figure}[tb]
\begin{center}
 \includegraphics[width=\columnwidth]{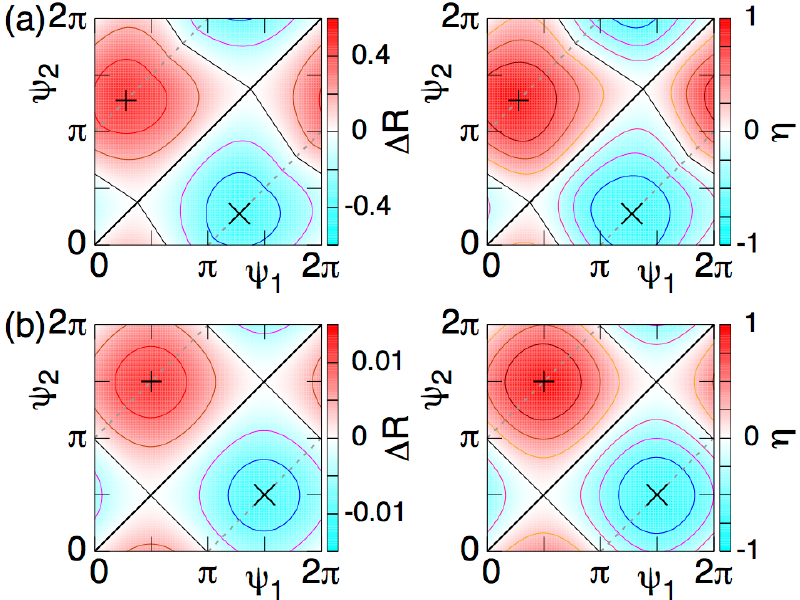}
\caption{(Colour online)
The net crawling displacement of the two-element cell over one period $\Delta R$ and the efficiency $\eta$ as functions of the phase shift $\psi_1$ and $\psi_2$ for different values of the intracellular dissipation rate: 
(a) $\xi=0.1$, and (b) $\xi=10$. 
The red (cyan) colour corresponds to the forward (backward) motion, i.e., the motion towards the $i=1$ (2) element. 
The phase shifts that give the maximum and the minimum of $\Delta R$ and $\eta$ are indicated by the plus and the cross signs, respectively. 
The diagonal thick black lines and the thin black lines represent the in-phase oscillatory motion and the reciprocating motion, neither of which show a net locomotion. 
On the grey dotted line, the two elements are in antiphase, i.e., $\psi_1$ and $\psi_2$ differ by $\pi$. 
}%
\label{fig:DeltaR_psi}
\end{center}
\end{figure}

\begin{figure}[tb]
\begin{center}
 \includegraphics[width=\columnwidth]{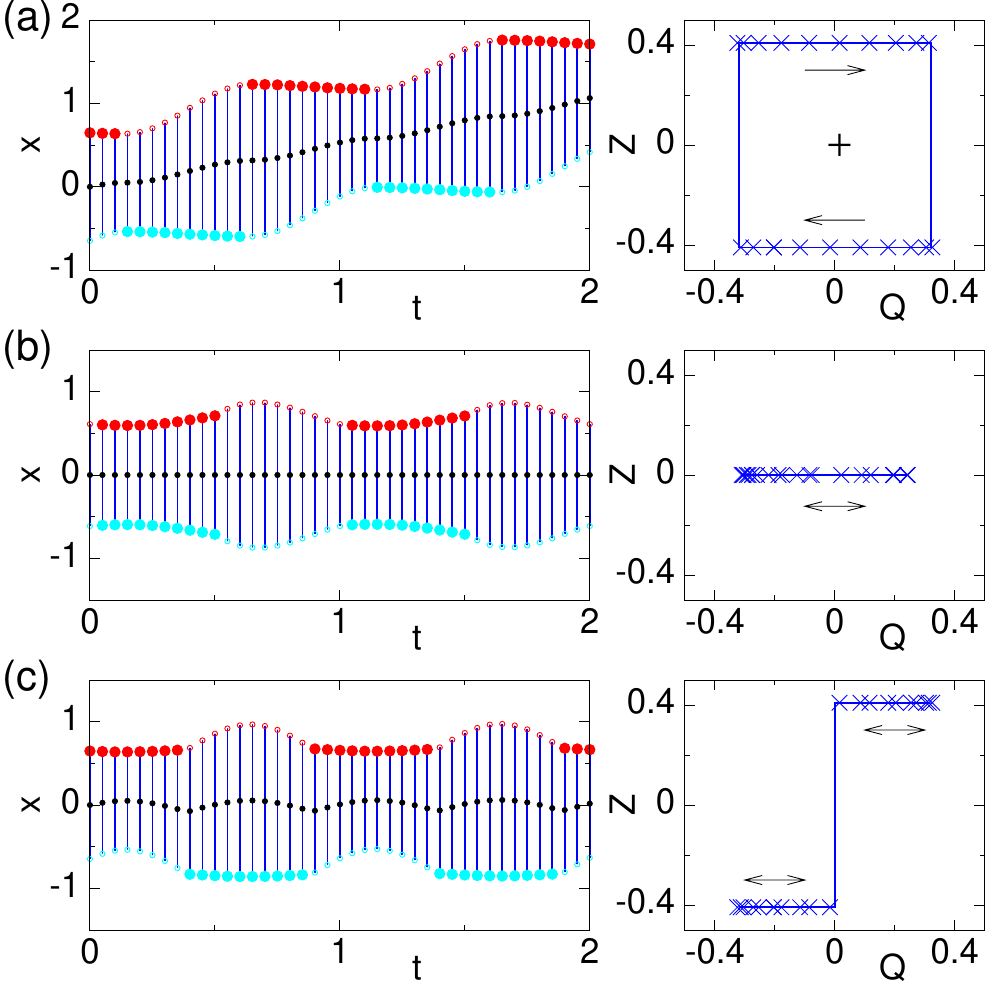}
\caption{(Colour online)
Two-element crawling cells for $\xi=0.1$ and 
(a) $\psi_1=0.27\pi$ and $\psi_2=1.27\pi$, 
(b) $\psi_1=\psi_2=\pi$, and
(c) $\psi_1=0.778794\pi$ and $\psi_2=1.778794\pi$. 
In each panel, the left subfigure shows the spatiotemporal plot for two periods.
The element 1 (2) is plotted by the red (cyan) circles, where the large and small ones correspond to the stick and slip states, respectively. 
The blue bars are the viscoelastic bond and the black dots represent the centre-of-mass position. 
The right subfigure displays the corresponding trajectory in the $Q$--$Z$ space. 
The blue crosses are obtained by numerical simulations, whereas the blue solid line is the analytic result. 
The arrows indicate the time evolution direction. 
}%
\label{fig:spatio_temporal}
\end{center}
\end{figure}

In fig.~\ref{fig:DeltaR_psi}, we plot the centre-of-mass displacement over one cycle $\Delta R$ and the efficiency $\eta$ for different values of the phase shift of the substrate friction $\psi_1$ and $\psi_2$. 
The intracellular dissipation rate is set as $\xi=0.1$ in fig.~\ref{fig:DeltaR_psi}(a) and $\xi=10$ in fig.~\ref{fig:DeltaR_psi}(b). 
We fixed the substrate friction coefficient of the stick and slip states as
\begin{equation}
\zeta_a=1,~\zeta_f=0.1. 
 \label{eq:value_zeta_a_f}
\end{equation}
The other parameters are set as $T=1$, $k=1$, $\ell_0=1.5$, and $\ell_1=0.5$. 
Due to the periodicity of the actuator and the substrate friction coefficient, $\Delta R$ and $\eta$ are $2\pi$ periodic in both $\psi_1$ and $\psi_2$. 
Note that, $\Delta R$ and $\eta$ are antisymmetric with respect to this diagonal line, since exchanging the phase shift, $\psi_1 \leftrightarrow \psi_2$, corresponds to switching the crawling direction. 
We also solved eqs.~\eqref{eq:force_balance_1} and \eqref{eq:force_balance_2} numerically for the same parameters and confirmed our analytical results. 
The spatiotemporal plots and the corresponding time series in the $Q$--$Z$ space are depicted in fig.~\ref{fig:spatio_temporal}. 

The maximum and minimum of $\Delta R$ and $\eta$ are indicated by the pluses and the crosses in fig.~\ref{fig:DeltaR_psi}, respectively. 
They correspond to the maximum migration in the forward and backward directions, respectively, and both of them occur when $\psi_1$ and $\psi_2$ are in antiphase, as depicted in fig.~\ref{fig:spatio_temporal}(a). 
Note that $\Delta R$ and $\eta$ are maximum (minimum) for the same values of the phase shift. 

When $\xi=10$, the maximum forward [backward] motion is achieved for $(\psi_1,\psi_2) = (\pi/2,3\pi/2)$ [$(3\pi/2,\pi/2)$], as shown in fig.~\ref{fig:DeltaR_psi}(b). 
This is because, when the intracellular dissipation rate $\xi$ is large and thus, the relaxation rate $\lambda$ is small, the first term on the right-hand side of eq.~\eqref{eq:dQ/dt} is negligible compared to the time derivative term on the left-hand side: 
\begin{equation}
\frac{d}{dt} Q(t) = \lambda \ell(t).
 \label{eq:dQ/dt_viscous}
\end{equation}
This means that the phase of the actual elongation $Q(t)$ is delayed by $\pi/2$ as compared to the actuator length $\ell(t)$. 
Therefore, the maximum forward [backward] motion occurs when the friction phase $\psi_1$ [$\psi_2$] is delayed by $\pi/2$ with respect to $\ell(t)$. 
Note that the displacement over one cycle is maximised when the substrate friction of one element, which eventually becomes the front of the cell, changes in phase with the actual elongation. 
Since the magnitude of the elongation is proportional to $\lambda$ as in eq.~\eqref{eq:dQ/dt_viscous}, the absolute value of the displacement $\Delta R$ decreases with $\lambda$. 

On the other hand, when $\xi$ is small and the relaxation rate $\lambda$ is large, the time derivative term on the left-hand side of eq.~\eqref{eq:dQ/dt} is negligible. 
Then, the actual elongation immediately adjusts to the actuator length: 
$Q(t) = \ell(t)$.
In this limit, the maximum forward [backward] displacement over one period occurs at $(\psi_1,\psi_2) = (0,\pi)$ [$(\pi,0)$]. 
Therefore, for the intermediate intracellular dissipation rate $\xi$, the optimum phase shift of the head element is found between 0 and $\pi/2$, as shown in fig.~\ref{fig:DeltaR_psi}(a). 
Here, note that, since $\zeta_i>0$, $\lambda$ does not increase infinitely large even for $\xi=0$. 

Finally, in each panel of fig.~\ref{fig:DeltaR_psi}, the thick and thin black lines correspond to the solutions where the cell cannot achieve a net locomotion, $\Delta R = 0$. 
The thick black line is a trivial case, where the substrate friction of the two elements are synchronized, $\psi_1 =\psi_2$. 
Therefore, over the course of the elongation and shrinking of the connecting bond, the centre-of-mass position does not move, around which the two elements oscillate symmetrically, as depicted in fig.~\ref{fig:spatio_temporal}(b). 
In contrast, on the thin black line, the cell exhibits reciprocating motion.
That is, it migrates in one direction for a certain time but then, for the rest of the period, moves in the opposite direction for the same distance. 
Therefore, after one period, it comes back to the original position and thus, does not achieve a net migration, as displayed in fig.~\ref{fig:spatio_temporal}(c). 

To summarise this letter, we have studied the mechanics of crawling cells on a substrate by using a minimal model that satisfies the force-free condition. 
The cell is modelled by two elements connected by a viscoelastic bond, representing the intracellular elasticity and dissipation, with an actuator that changes the length in time. 
The protrusion due to actin polymerization and actomyosin contraction are taken into account as the cyclic actuator elongation. 
The adhesion and deadhesion process are included in the periodic switch of the substrate friction between the stick and the slip states. 
Due to the phase shift of the adhesion-deadhesion transition of each element with respect to the actuator elongation, we have clarified that the time-reversal symmetry and thus, the translational symmetry, can be broken. 
In particular, the mismatch of the two phases affects the crawling efficiency and the direction of the locomotion. 

The maximum efficiency is realised when the substrate friction of the two elements change in antiphase. 
Their phase shift with respect to the actuator elongation also affects the efficiency. 
The optimum phase shift of the head element is found between 0 and $\pi/2$, which depends on the intracellular dissipation rate. 

In the most crucial case, if the two elements change the substrate friction in phase, the cell just oscillates its length with its centre of mass motionless. 
The cell also cannot achieve a net migration when it reciprocates and the centre of mass undergoes back-and-forth motion. 

We can straightforwardly extend our model to cells consisting of more subcellular elements, which may show richer dynamics. 
\begin{figure}[tb]
\begin{center}
 \includegraphics[width=\columnwidth]{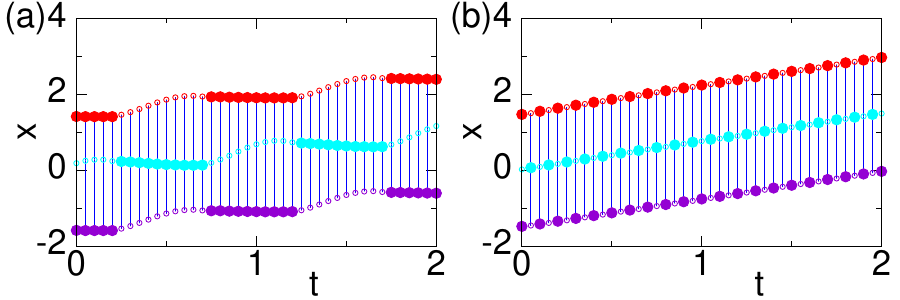}
\caption{(Colour online)
Three-element cells crawling with a time-independent length for different period of the internal cyclic motion: (a) $T=1$ and (b) $T=0.1$. 
The elements 1, 2, and 3 are plotted by the red, cyan, and purple circles, respectively. 
The phase shift is set as $\phi = \pi$, $\psi_1=\psi_3=0.4\pi$, and $\psi_2=1.4\pi$. 
}%
\label{fig:spatio_temporal_three_element}
\end{center}
\end{figure}
In reality, some cells such as keratocyte crawling on a substrate can migrate with a rather stationary shape~\cite{Keren2008Mechanism,Ohta2016Simple}. 
Such crawling motion with time-independent shape can be realised with our model if there are more than two elements. 
In fig.~\ref{fig:spatio_temporal_three_element}, we demonstrate it in the minimum case of a three-element cell. 
When the substrate friction coefficients of the outer two elements are synchronised but that of the middle one is in antiphase, and the actuator of the two bonds elongate in antiphase, the cell exhibits a locomotion without changing the total length, as shown in fig.~\ref{fig:spatio_temporal_three_element}(a). 
If the period of the elongation is much smaller than the typical time of the observation, 
the crawling motion becomes smooth, as displayed in fig.~\ref{fig:spatio_temporal_three_element}(b). 

Another possible extension of our model is to two dimensions~\cite{Tarama_Mori_Yamamoto}. 
In this case, in addition to a ballistic crawling motion, the cell can exhibit rotation. 
Moreover, we can further extend our current model to include intracellular chemical activities by reaction-diffusion equations~\cite{Tarama_Mori_Yamamoto}. 
Then, the elongation of the actuator, corresponding to the protrusion and contraction, as well as the adhesion-deadhesion transition to the substrate are induced by intracellular chemical reactions, which seems more realistic for actual cells. 

Finally, we mention that cells are often interacting not only with the substrate but also often with other cells. 
In fact, intercellular interactions are of great importance to multicellular dynamics such as wound healing and tissue formation. 
Such collective cellular dynamics was discussed in Ref.~\cite{Schnyder2017Collective} using a model basically corresponding to our current model in the special limit of $\psi_1=0$, $\psi_2=\pi$, $\zeta_a=\infty$, and $\zeta_f = 1$. 

Since our model satisfies the force-free condition, we expect that it will be the basis to investigate various physical phenomena in biology, especially those concerning mechanical properties such as mechanotaxis~\cite{Wang2006An,Kidoaki2010Mechanics}.
In addition, the present study will open new possibilities in new technologies such as designing biomedical micro-devices that can locomote by a simple actuation scheme in micro-scale complex environments such as biological tissues, where external forcing is difficult.

This work was supported by KAKENHI Grant No. 17H01083 from the Japan Society for the Promotion of Science (JSPS),  KAKENHI Grant No. 16H00765 from the Ministry of Education, Culture, Sports, Science, and Technology of Japan, and the JSPS Open Partnership Joint Research Projects.

\end{document}